\pdfoutput=1

\documentclass[
  aps,
  prb,
  twocolumn,
  superscriptaddress,
  nofootinbib
]{revtex4-2}

\usepackage{amsmath,amssymb}
\usepackage{graphicx}
\usepackage{hyperref}
\hypersetup{hidelinks}
\usepackage{microtype}

\newcommand{\me}{m_{\mathrm e}}
\newcommand{\kb}{k_{\mathrm B}}
\newcommand{\EF}{E_{\mathrm F}}
\newcommand{\Deltaind}{\Delta_{\mathrm{ind}}}
\newcommand{\gSm}{g_{\mathrm{Sm}}}
\newcommand{\gm}{g_{\mathrm m}}
\newcommand{\mN}{m_{\mathrm N}^{*}}
\newcommand{\msub}{m_{\mathrm{sub}}^{*}}
\newcommand{\Rsm}{R_{\mathrm s}}

\begin{document}

\title{Understanding the superconducting proximity effect in semiconductors through quantum oscillations}

\author{Milo Coombs}
\thanks{These authors contributed equally.}
\affiliation{Materials Department, University of California, Santa Barbara, California 93106, USA}

\author{Teun A. J. van Schijndel}
\thanks{These authors contributed equally.}
\affiliation{Department of Electrical and Computer Engineering, University of California, Santa Barbara, California 93106, USA}

\author{Yu Wu}
\affiliation{Department of Electrical and Computer Engineering, University of California, Santa Barbara, California 93106, USA}

\author{Jason T. Dong}
\affiliation{Materials Department, University of California, Santa Barbara, California 93106, USA}

\author{Yilmaz Gul}
\affiliation{London Centre for Nanotechnology, University College London, 17-19 Gordon Street, London WC1H 0AH, United Kingdom}

\author{Julian Choi}
\affiliation{Department of Physics, University of California, Santa Barbara, California 93106, USA}

\author{Christopher J. Palmstr{\o}m}
\email{cjpalm@ucsb.edu}
\affiliation{Materials Department, University of California, Santa Barbara, California 93106, USA}
\affiliation{Department of Electrical and Computer Engineering, University of California, Santa Barbara, California 93106, USA}

\author{Greg P. Mazur}
\email{gmazur@ucsb.edu}
\affiliation{Materials Department, University of California, Santa Barbara, California 93106, USA}
\affiliation{Department of Materials, University of Oxford, Parks Road, Oxford OX1 3PH, United Kingdom}

\begin{abstract}
Superconductor--semiconductor hybrids host emergent states of matter and offer a platform for new qubits, but the superconducting metal shunts electrical transport, which rules out conventional semiconductor characterization and leaves the hybrid parameters to speculation. Here we determine density, mass, $g$-factor, mobility and subband occupation beneath the superconductor, from Shubnikov--de Haas oscillations of a buried InAs quantum well under Al, Sn, V, Nb, Ta and Re films, with a Dingle analysis that accounts for the shunt. Every metal adds an interface subband whose occupation falls into one of two classes, whereas the mass and $g$-factor of the buried well are unchanged to within 10\%. Within the uncertainty set by the transport mobility, no film shortens the quantum lifetime of the buried well, and Al and Sn lengthen it. Quantum lifetimes bound the hybridization of the interface subband to 2--4~meV. These measurements supply the normal-state parameters that tunnelling spectroscopy renormalizes but cannot measure.
\end{abstract}

\maketitle

When a semiconductor is coupled to a superconductor, the hybrid is expected to inherit properties of both. This is the basis of engineered topological superconductivity~\cite{FuKane2008,Lutchyn2010,Oreg2010}, gate-tunable superconducting electronics and hybrid quantum devices. Epitaxial growth has produced hard induced gaps and a growing range of parent superconductors~\cite{Krogstrup2015,Shabani2016,Chang2015,Carrad2020,Pendharkar2021,Mazur2022,Kanne2021}, enabling Andreev and gatemon qubits~\cite{Hays2021,Larsen2015,deLange2015}, Cooper-pair splitters~\cite{Hofstetter2009,Wang2022} and minimal Kitaev chains~\cite{Dvir2023}. The superconductor is not a passive pairing reservoir. Hybridization is expected to renormalize the semiconductor dispersion, Zeeman response and spin--orbit coupling by amounts set by the interface transparency and the parent gap~\cite{Reeg2018,Antipov2018,Mikkelsen2018}, while disorder remains a central obstacle to topological protection~\cite{Ahn2021,Aghaee2023}.

In experiments, the parameters that enter models of the superconducting state are almost never measured in the hybrid itself. The metal shunts the semiconductor, so density, mobility, effective mass and $g$-factor are inferred from etched Hall bars, uncapped sister wafers or device spectroscopy. Each is a physically different system. Removing the metal changes the electrostatic boundary condition, and etching modifies the surface that forms the accumulation layer~\cite{Yuan2020}. Yet it remains unknown how far parameters measured on such structures describe the semiconductor that is actually coupled to the superconductor. Hybrid parameters have been left to educated guesswork, which makes simulations of hybrid devices speculative.

In this article, we measure the semiconductor beneath the superconductor directly, through the Shubnikov--de Haas (SdH) oscillations of a buried InAs quantum well under films of Al, Sn, V, Nb, Ta and Re. In a disordered metal film the Landau levels are unresolved and the magnetoresistance is smooth. In the two-dimensional electron gas (2DEG) the density of states at the Fermi level oscillates with $1/B$, and with it the scattering rate and the conductance of each subband. The oscillatory part of the measured resistance therefore comes from the semiconductor alone, even when the metal carries most of the current, as in topological-insulator films~\cite{Qu2010,Analytis2010,Dybko2017} and recent hybrid heterostructures~\cite{Zimmerman2026}. Its frequencies, and their temperature and tilt dependence, are those of the semiconductor subbands. Its amplitude is not, because the shunt filters it, which is where conventional analysis fails. We make four observations. Every metal creates an interface subband and pins it at one of two levels that do not follow the work function. The cyclotron mass and $g$-factor of the buried well are the same beneath all six metals as in the uncapped well. No film degrades the quantum lifetime of the buried well, once the filtering by the shunt is removed from the amplitude, and the two $sp$ metals improve it. The buried quantum well couples to the metal far more weakly than the interface subband. Tunnelling spectroscopy of a hybrid measures the induced gap and the subgap $g$-factor but not the density, mass or normal-state $g$-factor that those two numbers renormalize. Quantum oscillations supply exactly those, so the two techniques together constrain the low-energy parameters of the semiconductor/superconductor hybrids.

\begin{figure*}[t]
\centering
\includegraphics[width=\textwidth]{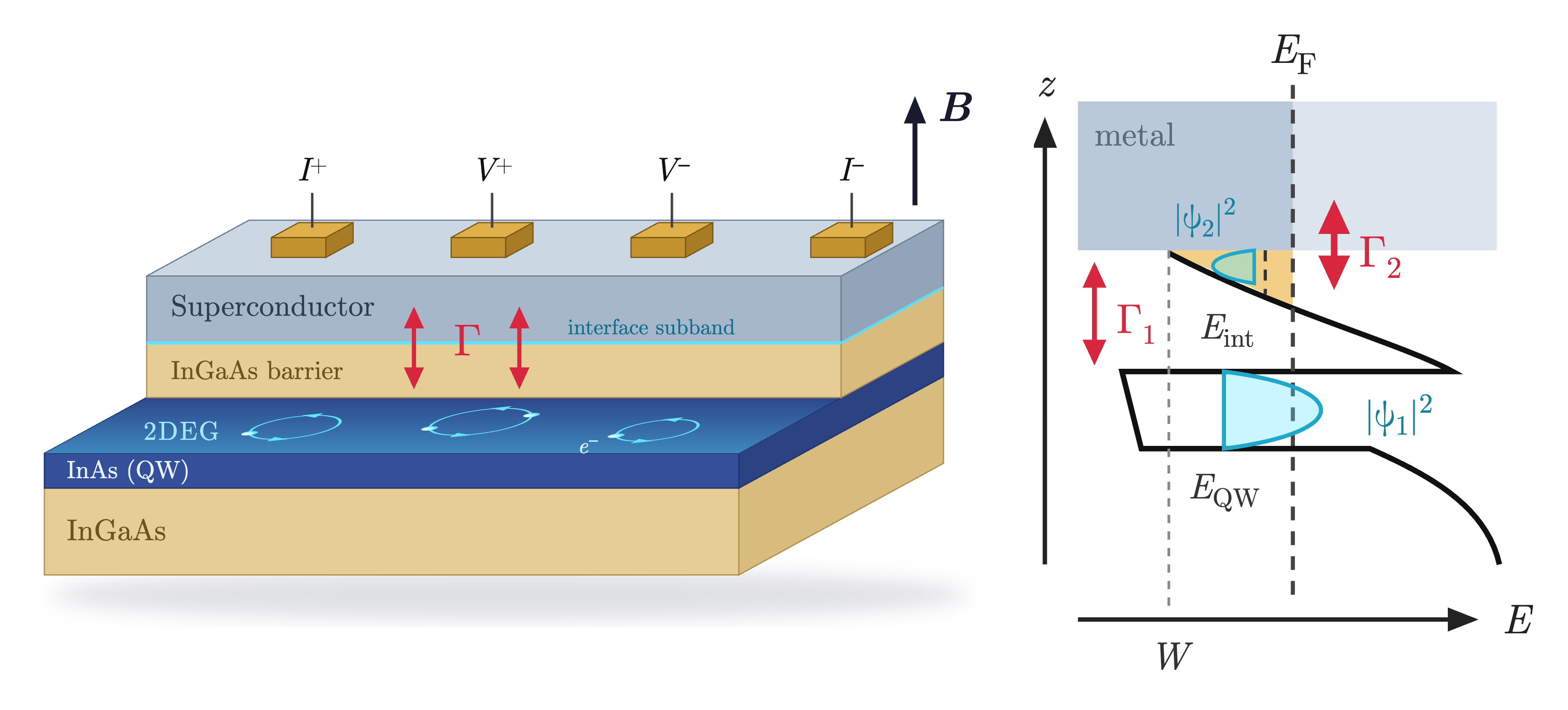}
\caption{\textbf{Quantum-oscillation spectroscopy of a buried quantum well beneath a superconducting film.}
\textbf{Left}, heterostructure and in-line four-probe geometry. A buried InAs quantum well lies below an InGaAs barrier and the metal film. SdH oscillations of the semiconductor subbands are measured on unpatterned material in a magnetic field $B$.
\textbf{Right}, schematic conduction-band profile along the growth direction $z$, not to scale. The occupied states comprise a fundamental state localized in the buried InAs quantum well, $|\psi_{\mathrm{QW}}|^2$, and a metal-induced accumulation state localized near the metal/InGaAs interface, $|\psi_{\mathrm{int}}|^2$. Their band minima are $E_{\mathrm{QW}}$ and $E_{\mathrm{int}}$, and the Fermi energy is $\EF$. The states couple to the metal with rates $\Gamma_{\mathrm{QW}}$ and $\Gamma_{\mathrm{int}}$. $W$ is the InAs quantum-well width.}
\label{fig:setup}
\end{figure*}

The heterostructure is an In$_{0.75}$Ga$_{0.25}$As (10~nm)/InAs (7~nm)/In$_{0.75}$Ga$_{0.25}$As quantum well (Fig.~\ref{fig:setup}). We refer to them as B1 and B2 throughout the manuscript. The films were deposited by electron-beam evaporation (Ta,Nb,Re) or through effusion cell (Al,Sn,V) at 7~K and protected by a nominal 2.5-nm Al layer oxidized in situ before any warm-up, so the cap reaches room temperature as an oxide and forms no superconducting parallel layer (Methods). Four-terminal magnetotransport was measured on unpatterned material with collinear probes.

\begin{figure*}[ht!]
\centering
\includegraphics[width=\textwidth]{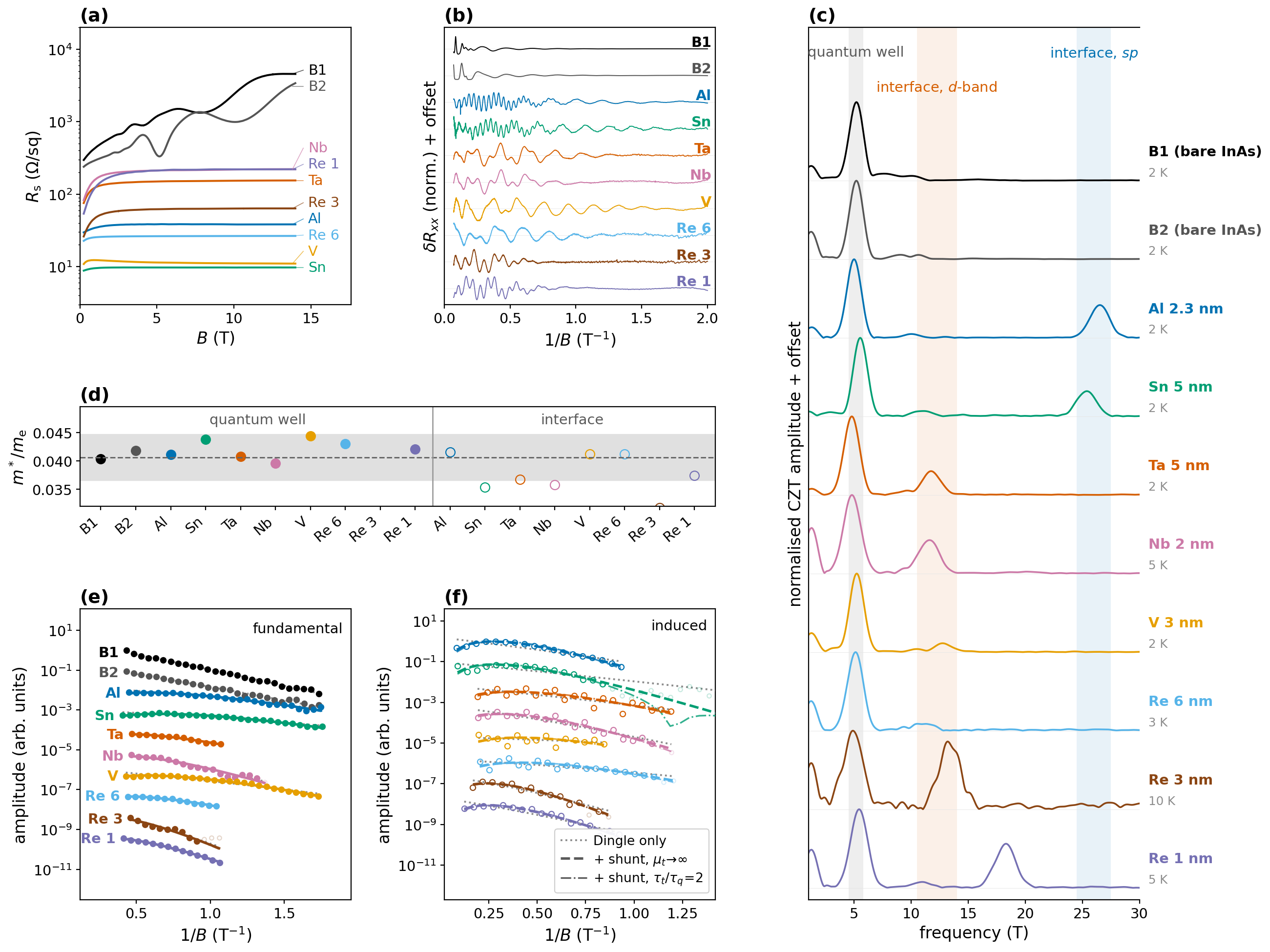}
\caption{\textbf{Quantum oscillations of the buried well beneath six superconducting films.}
Colours identify the sample in every panel. B1 and B2 are the two uncapped reference wells.
\textbf{a}, Smooth sheet resistance of each stack. The uncapped wells are one to three decades more resistive than the metallized ones, and this ratio sets the size of the amplitude correction in \textbf{e} and \textbf{f}.
\textbf{b}, Background-subtracted oscillations against inverse field at the lowest normal-state temperature, normalized and offset.
\textbf{c}, Chirp-$z$ spectra of the traces in \textbf{b}, with the analysis temperature beside each. The quantum-well frequency stays near 5~T in all ten samples, whereas metallization adds a second branch at $11$--$13.4$~T beneath the $d$-band metals and $25$--$27$~T beneath Al and Sn.
\textbf{d}, Cyclotron masses of all branches, quantum-well branches at left and interface branches at right, filled and open symbols respectively. Dashed line and shading, the quantum-well mean and $\pm10\%$. The 3-nm Re sample has no mass because its film is normal over the fit window only above 10~K.
\textbf{e}, Oscillation amplitude of each fundamental branch against $1/B$, normalized and offset by decades. Grey dotted curves are conventional Dingle fits. Coloured curves are fits of Eq.~\eqref{eq_shunt_model} at $\mu_t\to\infty$ (solid) and at $\tau_t/\tau_q=2$ (dash-dot). The uncapped wells need no correction.
\textbf{f}, The same for the interface branches, in the order and colours of \textbf{e}. These envelopes are non-monotonic. Faint open symbols lie outside the fit window. The resulting mobilities are listed in Table~\ref{tab_summary}.}
\label{fig:oscillations}
\end{figure*}

The uncapped well hosts a single subband at $F_0=5.2$~T, corresponding to $n=2.5\times10^{11}$~cm$^{-2}$ (Fig.~\ref{fig:oscillations}c). Depositing a metal leaves this oscillation intact and adds a second frequency that is absent from the bare well. Its value falls into two classes. Under Ta, Nb, V and Re it lies at $11.0$--$13.4$~T, corresponding to $(5.3$--$6.5)\times10^{11}$~cm$^{-2}$. Under Al and Sn it lies at $25$--$27$~T and carries $(1.2$--$1.3)\times10^{12}$~cm$^{-2}$, five times the density of the well. A Re thickness series locates the new state within a few nanometres of the film. For 6- and 3-nm Re it remains in the refractory class, at $11.0$ and $13.4$~T, whereas at 1~nm it moves to $18.3$~T, and the quantum-well frequency does not move at all. This suggest that a state whose envelope sits at the metal/InGaAs boundary can respond to a boundary condition a few nanometres above it while the well 10~nm below does not.

\begin{figure*}[ht]
\centering
\includegraphics[width=\textwidth]{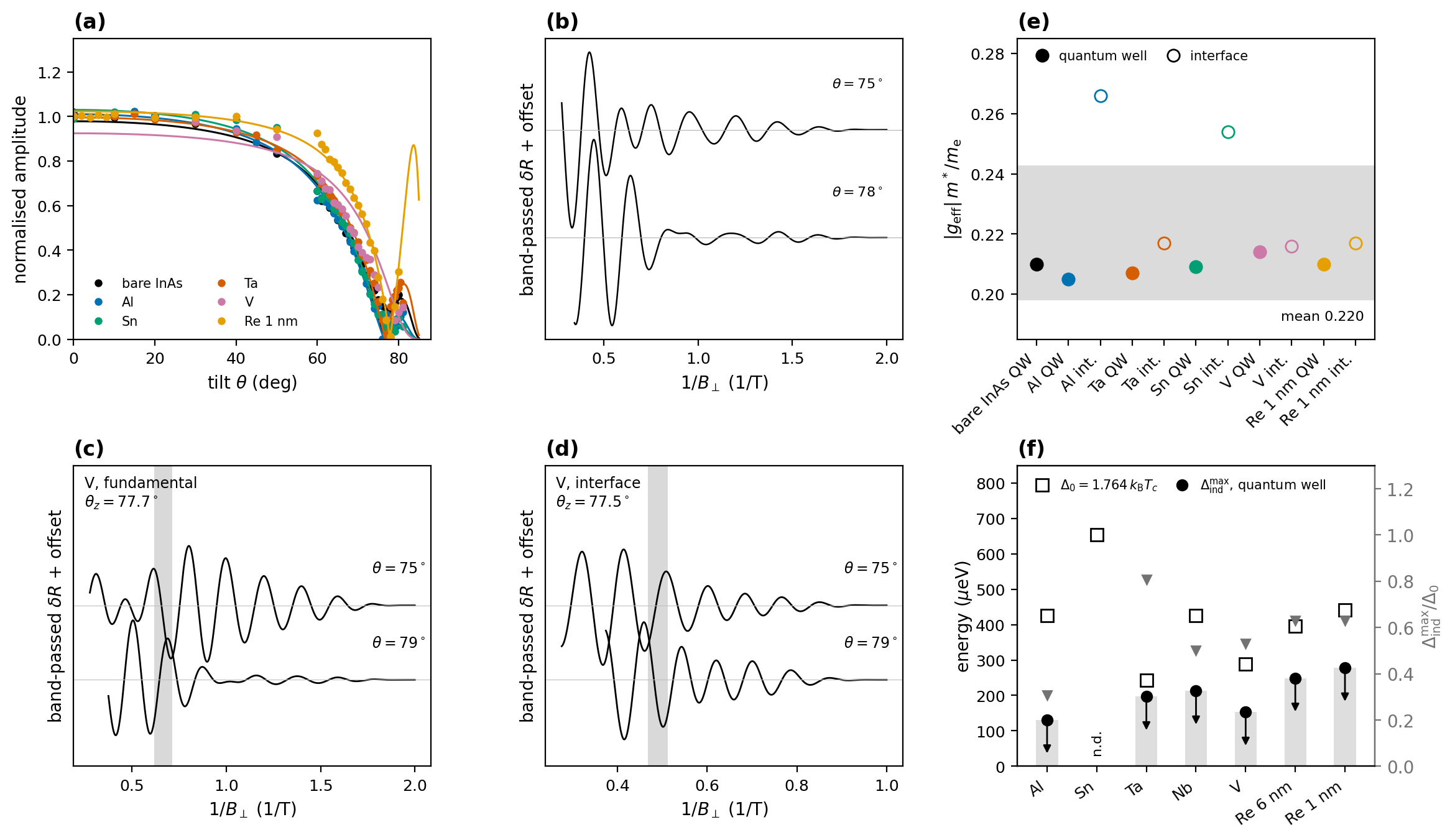}
\caption{\textbf{Spin-zero spectroscopy and the induced-gap ceiling of the buried well.}
\textbf{a}, Tilt dependence of the fundamental-branch amplitude for B2 and the Al-, Sn-, Ta-, V- and 1-nm-Re-capped samples, normalized to the low-angle value. Curves are damped coincidence fits. All six pass through their first spin zero near $78^\circ$.
\textbf{b}, Band-passed oscillations of B2 either side of $\theta_z$, offset vertically. The inversion identifies a spin zero without a lineshape model. Shading marks half a period.
\textbf{c,d}, The same for the V fundamental (\textbf{c}) and V interface (\textbf{d}) branches at $75^\circ$ and $79^\circ$.
\textbf{e}, Spin-zero products $|g_{\mathrm{eff}}|m^*/\me=\cos\theta_z$ for all eleven branches, filled for quantum-well and open for interface states, coloured as in \textbf{a}. Dashed line, mean 0.220. Shading, $\pm10\%$. Every quantum-well branch lies within $0.205$--$0.214$. The Al and Sn interface branches return $0.266$ and $0.254$, the two cases where $\theta_z$ falls near the edge of the measured range.
\textbf{f}, Upper limits on the gap induced in the buried well, from Eq.~\eqref{eq_gamma_differential} through the pole of Eq.~\eqref{eq_selfenergy}. They are what the lifetimes exclude. Filled circles with arrows and shaded bars, $\Deltaind^{\max}$. Open squares, the parent gap $\Delta_0=1.764\,\kb T_c$. Grey triangles, the ratio (right axis), $0.31$--$0.81$. Sn has no bound (n.d.) because its buried-well lifetime exceeds that of both uncapped samples.}
\label{fig:spinzero}
\end{figure*}

Based on the above observation, we argue that the new branch is not an excited orbital of the well. Its occupation energy is $33$--$40$~meV below $\EF$ for the refractory metals and $85$--$99$~meV for Al and Sn, against $14$--$16$~meV for the fundamental state (Methods, Eq.~\eqref{eq_subband_occupation}), so its band minimum lies $18$--$26$~meV (refractory metals) or $70$--$84$~meV (Al and Sn) below that of the fundamental state, which an excited orbital cannot do. It is not a harmonic of the fundamental oscillation, although $2F_0=9.6$--$11.0$~T lies close to the refractory branch. A second harmonic fitted with the Lifshitz--Kosevich form returns twice the mass and places its first spin zero near $64^\circ$, whereas the measured interface masses are $0.035$--$0.042\,\me$, equal to the fundamental mass within 20\%, and the Ta and V interface branches pass through their spin zero near $77^\circ$ like every other branch (Fig.~\ref{fig:spinzero}). The frequency of every interface branch measured in tilt, six in all, scales as $F(0)/\cos\theta$, as required for a two-dimensional orbit, which rules out any origin in the film or the cap. We therefore refer to it as the metal-induced interface subband.

The two classes reflect the boundary condition at the interface, and this can be stated without converting an occupation into a barrier height. Al and Ta differ by 30~meV in work function~\cite{Michaelson1977} but by 45~meV in the occupation energy of the interface subband, which places them in opposite classes, whereas 6-nm Re and V differ by 660~meV in work function and by 5~meV in occupation, which places them in the same class. Across the four refractory metals the occupation is uncorrelated with the work function. The interface is therefore strongly pinned, as is well established for InAs surfaces and contacts~\cite{Tung2001,Noguchi1991,Feng2016,Schuwalow2021}, but at two distinct levels, so the pinned level is not universal. One property that divides the set as observed is the orbital character of the metal states at the Fermi level, $sp$-like for Al and Sn and partially filled $d$ bands for V, Nb, Ta and Re, and in the metal-induced-gap-state picture the interface spectrum inherits that character~\cite{Heine1965,Tersoff1984,Louie1977,Sankey1984}. Interface chemistry could shift the neutrality level as effectively, although the 7~K deposition freezes interdiffusion kinetically, and cross-sectional electron microscopy of the samples studied here shows abrupt interfaces without a reaction layer~\cite{TEMref}. Film thickness and crystalline phase are not the deciding factors either. Al at 2.3~nm (10 monolayers) and Nb at 2~nm fall in opposite classes, as do Sn and Ta at 5~nm, while crystalline 6-nm Re and amorphous 3-nm Re fall in the same one.

Thermal-expansion mismatch also divides the set. Between 300 and 4~K, Al and Sn contract by 0.3 \% more than InAs in linear dimension~\cite{Corruccini1961}, they are far more ductile than the refractory metals, and every sample is warmed to room temperature and cooled again before measurement. A band shift frozen in by plastic relaxation of the film nevertheless fails on two counts. Force balance for a nanometre-thick film on a thick substrate leaves the substrate surface strained only by wafer curvature, $\varepsilon\sim10^{-7}$~\cite{Freund2003}, whereas the 45~meV separating the classes requires an in-plane strain of $\varepsilon=8\times10^{-3}$~\cite{Vurgaftman2001}. A film that relaxes at room temperature is in tension when cold, which places the substrate surface in compression and raises the InAs band edge, so relaxation would make the Al and Sn wells shallower, whereas they are the deepest in the series. The Al interface frequency also reproduces to within $0.3\%$ of itself between cooldowns (SI).

We next examine effective masses in the samples under study. The Lifshitz--Kosevich mass of the fundamental branch is $0.040$--$0.044\,\me$ under every metal, within 10\% of the bare value (Fig.~\ref{fig:oscillations}d) and above the InAs band-edge mass as expected from nonparabolicity~\cite{Yuan2020}. In a tilted field the spin factor of the oscillation crosses zero when the Zeeman energy reaches half the cyclotron spacing, $|g_{\mathrm{eff}}|m^*/\me=\cos\theta_z$, Eq.~\eqref{eq_spinzero}, which fixes $|g|m^*$ without assuming the mass (Methods). Every fundamental branch passes through its first spin zero near $78^\circ$ (Fig.~\ref{fig:spinzero}a), the waveforms invert across the null (Fig.~\ref{fig:spinzero}b--d), and the products fall within $0.205$--$0.214$ for all six quantum-well branches (Fig.~\ref{fig:spinzero}e). With the measured masses this gives a common $|g_{\mathrm{eff}}|=4.9\pm0.2$ for the buried well beneath every film. The three refractory interface branches give $|g_{\mathrm{eff}}|=5.3$--$5.9$. The Al and Sn interface branches are unresolved, their nulls falling near the edge of the measured tilt range, and we do not quote values for them. These invariances follow from the form of the tunnelling self-energy. In the normal state its wide-band limit is purely imaginary, which broadens the Landau levels but cannot shift the cyclotron spacing or the Zeeman-to-cyclotron ratio (Methods). Corrections beyond that limit, from coherent hybridization or an energy-dependent metal density of states~\cite{Antipov2018,Mikkelsen2018}, could shift both. We resolve no such shift at the 10\% level. If the Landau-quantized state at the Fermi level carried an energy-independent fraction $w$ of its weight in the metal, its inverse mass and $g$-factor would be weighted averages of the semiconductor and metal values, and the observed invariance bounds $w\lesssim0.1$, for any metal mass much larger than $0.04\,\me$ and within this two-component model. This concerns the long-lived quasiparticles that produce the oscillations. Weight broadened beyond the oscillation linewidth is invisible here, and both are distinct from the dynamical weight $1-Z$ that pairing generates below the gap.

Whether metallization changes the disorder depends on the film. Beneath a film the oscillation amplitude is not proportional to the Dingle factor of the semiconductor branch. Channels add in conductivity, so the measured resistance responds to the oscillatory conductivity of the subband, $\delta R=-(\Rsm^2/g)\,\delta\sigma_{xx}$, where $\Rsm(B)$ is the smooth sheet resistance of the stack and $g$ the geometric factor, and $\delta\sigma_{xx}$ carries the Drude prefactor $P=(\mu_t^2B^2-1)/(1+\mu_t^2B^2)^2$ of the branch's transport mobility $\mu_t$. For an uncapped well the tensor inversion cancels this prefactor exactly and the textbook Dingle plot applies~\cite{Ando1982,Coleridge1989,Coleridge1991,Candido2023}. For a shunted well it does not. The logarithmic slope of $P$ over our field windows is comparable to the Dingle slope itself, and it is the origin of the non-monotonic envelopes of every metallized branch in Fig.~\ref{fig:oscillations}e,f, which no Dingle exponential can produce. We therefore fit each branch to
\begin{equation}
A(x)\propto\frac{X}{\sinh X}\;\Rsm^2(B)\;\bigl|P(B;\mu_t)\bigr|\;e^{-\pi x/\mu_q},
\label{eq_shunt_model}
\end{equation}
with $x=1/B$ and $X/\sinh X$ the thermal factor evaluated with the branch's own mass (Methods). The smooth background does not determine $\mu_t$ in a two-subband hybrid (Methods), so we quote $\mu_q$ as an interval. Its lower end is the $\mu_t\rightarrow\infty$ limit, in which the fit has no free parameter beyond $\mu_q$. Its upper end is the value at $\tau_t/\tau_q=2$, the smallest ratio measured on the uncapped wells from their zero-field resistance, which screening and isotropic escape can only reduce. The interval reproduces to $1$--$10\%$ across analysis temperatures of $2$--$6$~K (SI), whereas a conventional Dingle plot of the same data moves by factors of $1.5$--$4$ with the analysis temperature.

With this correction no film shortens the quantum lifetime of the buried well within the uncertainty set by the transport mobility. Every interval overlaps or exceeds that of B1, and the Sn and Al intervals also overlap or exceed that of B2 (Table~\ref{tab_summary}). The two classes of metal differ. The lower bound is $(1.1$--$1.2)\times10^4$~cm$^2$V$^{-1}$s$^{-1}$ beneath Al and Sn against $(0.8$--$1.0)\times10^4$ beneath Ta, V and Re, an ordering of the bounds that does not depend on the transport mobility, although the intervals themselves overlap at their upper ends. Beneath Sn the well is at least as clean as either reference at the lower end of its interval and up to 30\% (B2) or 60\% (B1) cleaner at the upper end. The improvement correlates with the interface subband. The $sp$ metals hold twice the interface density of the $d$-band metals, $(1.2$--$1.3)\times10^{12}$ against $(5.3$--$6.5)\times10^{11}$~cm$^{-2}$, and their interface branches are the least mobile in the series, $(4$--$6)\times10^3$ against $(5$--$9)\times10^3$~cm$^2$V$^{-1}$s$^{-1}$. A dense interface layer screens the remote charge that limits the buried well. Smoother boundaries beneath the $sp$ metals cannot be excluded, but thickness is not the variable, since Al at 2.3~nm and Nb at 2~nm fall in opposite classes. We note that the V interface branch is the weakest in the series and its envelope does not follow Eq.~\eqref{eq_shunt_model} over any usable window, so we quote no mobility for it.

We now use the lifetimes to bound the coupling. In the tunnelling-Hamiltonian description, escape into the metal is a golden-rule process~\cite{Bardeen1961}, its rate adds to the disorder rate in the Landau-level width, and the disorder rate cannot be negative, so $\Gamma\leq\hbar/2\tau_q$ for each branch (Methods). For the interface subbands this gives $\Gamma_{\mathrm{int}}\leq1.9$--$3.2$~meV, three to ten times the parent gaps $\Delta_0=1.764\,\kb T_c=0.24$--$0.85$~meV, so their lifetimes limit the induced gap only to within $2$--$13\%$ of $\Delta_0$ (Table~\ref{tab_summary}). A triangular-well estimate of the attempt frequency translates the bound into a transmission of at most $0.06$--$0.2$ per encounter with the boundary (Methods), and a transmission of a few percent is already enough to place these states in the regime $\Gamma\gtrsim\Delta_0$.

The buried well admits a sharper bound, because the uncapped samples measure its disorder rate with no metal present. Subtracting the two rates, and assuming only that the film does not reduce the disorder, gives $\Gamma_{\mathrm{QW}}\leq0.18$~meV beneath Al, $0.28$ beneath V, $0.31$ beneath Nb and $0.5$--$0.6$ beneath Ta and Re (Methods), and induced-gap ceilings of $0.3$--$0.8\,\Delta_0$ (Fig.~\ref{fig:spinzero}f). The assumption fails for Sn, whose buried-well lifetime exceeds that of both references, so there the film has evidently reduced the disorder and we quote no bound. If a film also reduces the disorder, as Sn evidently does and as the screening mechanism above implies for Al, an escape rate comparable to that reduction is masked, so the ceilings of Fig.~\ref{fig:spinzero}f hold only where the disorder is unchanged and are the constraint the normal-state data place on the coupling rather than a measurement of it. The lifetime of a well in direct contact with Al is measurably shortened by escape into the film~\cite{Zimmerman2026}, and no such shortening is resolved here. Beyond that, the normal-state data do not separate the escape rate of the buried well from its disorder rate, and the ceilings above are the constraint they place on it.

These normal-state parameters are what the superconducting state renormalizes, and they are exactly the parameters that tunnelling spectroscopy cannot supply. Tunnelling into a hybrid measures the induced gap and, from the Zeeman splitting of subgap states, the subgap $g$-factor~\cite{Vaitiekenas2018,vanLoo2023}, but neither the density nor the mass nor the normal-state $g$-factor from which those values descend. Integrating out the superconductor gives a quasiparticle weight $Z=(1+\Gamma/\Delta_0)^{-1}$ that acts on every semiconductor term in the same way. It enhances the mass, $\msub=\mN/Z$, interpolates the $g$-factor between the semiconductor and metal values, $g_{\mathrm{sub}}=Z\gSm+(1-Z)\gm$, and reduces the spin--orbit coupling, $\alpha_{\mathrm{sub}}=Z\alpha$ (Methods, Eqs.~\eqref{eq_Z}, \eqref{eq_renorm} and \eqref{eq_gapZ})~\cite{Reeg2018,Antipov2018}. $Z$ runs from 1 for an uncoupled semiconductor towards 0 for a state absorbed into the metal, and the measured subgap $g$-factor tests whether one $Z$ accounts for both the gap and the splitting. Where a beat in the oscillations resolves the Rashba coupling, as in Ref.~\cite{Zimmerman2026}, the same $Z$ fixes $\alpha_{\mathrm{sub}}$, so that quantum oscillations and tunnelling spectroscopy together constrain the low-energy Hamiltonian of Eqs.~\eqref{eq_Z}--\eqref{eq_gapZ}. Our lifetimes constrain $Z$ only to $Z\geq0.1$--$0.2$ depending on the film, because they allow $\Gamma$ up to three to ten times $\Delta_0$.  For an induced gap of $0.5\,\Delta_0$, typical of Al on InAs~\cite{Shabani2016,Aghaee2023}, the pole of Eq.~\eqref{eq_selfenergy} gives $\Gamma=0.9\,\Delta_0$ and $Z=0.54$. An interface subband with the normal-state values measured here, $\mN=0.04\,\me$ and $\gSm=-5.5$, then acquires $\msub=0.08\,\me$ and, with $\gm=+2$ for the metal, $g_{\mathrm{sub}}=-2.0$, a nearly threefold suppression of the Zeeman splitting.

\paragraph*{Conclusion} In summary, quantum oscillations measured beneath six superconductors determine the normal-state parameters of the semiconductor in the intact hybrid. Every metal creates an interface subband, whose occupation falls into one of two classes across this set of six and which the work function does not predict. The buried well retains its mass, its $g$-factor and its quantum lifetime beneath every film, and the $sp$ metals lengthen that lifetime. Three consequences follow for devices. The as-grown stack carries two subbands, and the measured band offsets set the depletion required to reach the odd-channel regime of the simplest Majorana constructions~\cite{Lutchyn2010,Oreg2010} and the coupling budget of quantum-dot Kitaev chains built on such stacks~\cite{Dvir2023,tenHaaf2024,Bordin2025}. A superconducting film is a pairing reservoir that does not by itself broaden the well, so where the disorder is unchanged the coupling and the level width can be chosen separately. All of these quantities are measured on unpatterned material. These conclusions apply to the In$_{0.75}$Ga$_{0.25}$As-barrier stacks that are the most widely used platform for InAs two-dimensional electron gases, and they are not universal properties of InAs hybrids. The influence of a metal is a property of the complete metal-barrier-well stack. The top barrier sets the wavefunction amplitude at the metal, and with it the interface occupation, the screening and the hybridization rate, as Zimmerman \textit{et al.} showed by varying the Al content of In$_{1-x}$Al$_x$As barriers beneath a fixed Al film~\cite{Zimmerman2026}. The same metal produces different subband occupations and coupling scales behind different barriers, and the method reported here measures each of them in the stack in which it is used, on any platform for which quantum oscillations survive metallization, including Ge/SiGe and van der Waals stacks.

\begin{table*}[t]
\caption{\textbf{Measured subband parameters and coupling bounds.}
Rows are grouped by subband, and film thicknesses are given with the film. Densities follow from $n=2eF/h$. Quantum mobilities are intervals from Eq.~\eqref{eq_shunt_model} at the lowest normal-state temperature, from the $\mu_t\rightarrow\infty$ limit to the $\tau_t/\tau_q=2$ ceiling. $\Gamma^{\max}$ is $\hbar/2\tau_q$ for the interface subbands, Eq.~\eqref{eq_gamma_bound}, and the differential bound of Eq.~\eqref{eq_gamma_differential} for the quantum-well branches, both evaluated at the shortest lifetime. $\Deltaind^{\max}$ follows from the pole of Eq.~\eqref{eq_selfenergy} with $\Delta_0=1.764\,\kb T_c$, which for the strong-coupling films Nb, Ta and V underestimates the gap by up to 20\%. All coupling entries are upper limits set by the lifetimes. Entries marked n.d.\ are not determined (Methods). The 3-nm Re interface lifetime assumes $m^*=0.042\,\me$.}
\label{tab_summary}
\centering
\footnotesize
\begin{ruledtabular}
\begin{tabular}{llcccccccc}
subband & film & $T_c$ & $F$ & $n$ & $m^*/\me$ & $\mu_q$ & $\tau_q$ & $\Gamma^{\max}$ & $\Deltaind^{\max}$ \\
 & & (K) & (T) & ($10^{11}$\,cm$^{-2}$) & & ($10^{3}$\,cm$^2$/Vs) & (ps) & (meV) & ($\mu$eV) \\
\hline
quantum & B1 & n/a & 5.2 & 2.5 & 0.040 & 9.9 & 0.23 & n/a & n/a \\
well & B2 & n/a & 5.2 & 2.5 & 0.042 & 12.1 & 0.28 & n/a & n/a \\
 & Al, 2.3 nm & 2.8 & 5.0 & 2.4 & 0.041 & 10.9--16.0 & 0.25--0.37 & 0.18 & 131 \\
 & Sn, 5 nm & 4.3 & 5.5 & 2.7 & 0.044 & 12.2--15.9 & 0.31--0.40 & n.d. & n.d. \\
 & Ta, 5 nm & 1.6 & 4.8 & 2.3 & 0.041 & 8.4--12.3 & 0.20--0.29 & 0.56 & 193 \\
 & Nb, 2 nm & 2.8 & 4.8 & 2.3 & 0.040 & 10.2--15.4 & 0.23--0.35 & 0.31 & 195 \\
 & V, 3 nm & 1.9 & 5.2 & 2.5 & 0.044 & 9.3--12.6 & 0.23--0.32 & 0.28 & 153 \\
 & Re (cryst.), 6 nm & 2.6 & 5.1 & 2.5 & 0.043 & 8.1--10.8 & 0.20--0.26 & 0.52 & 249 \\
 & Re (amorph.), 3 nm & 5.6 & 4.9 & 2.4 & n.d. & n.d. & n.d. & n.d. & n.d. \\
 & Re (amorph.), 1 nm & 2.9 & 5.4 & 2.6 & 0.042 & 8.4--12.8 & 0.20--0.31 & 0.55 & 273 \\
\hline
interface & Al, 2.3 nm & 2.8 & 26.5 & 12.8 & 0.042 & 4.3--5.9 & 0.10--0.14 & 3.2 & 412 \\
 & Sn, 5 nm & 4.3 & 25.4 & 12.3 & 0.035 & 4.3--5.8 & 0.09--0.12 & 3.8 & 619 \\
 & Ta, 5 nm & 1.6 & 11.7 & 5.7 & 0.037 & 6.5--9.5 & 0.14--0.20 & 2.4 & 238 \\
 & Nb, 2 nm & 2.8 & 11.6 & 5.6 & 0.036 & 6.0--9.2 & 0.12--0.19 & 2.7 & 407 \\
 & V, 3 nm & 1.9 & 12.4 & 6.0 & 0.041 & n.d. & n.d. & n.d. & n.d. \\
 & Re (cryst.), 6 nm & 2.6 & 11.0 & 5.3 & 0.042 & 7.1--9.2 & 0.17--0.22 & 1.9 & 368 \\
 & Re (amorph.), 3 nm & 5.6 & 13.4 & 6.5 & 0.042 & 5.0--6.8 & 0.12--0.16 & 2.7 & 739 \\
 & Re (amorph.), 1 nm & 2.9 & 18.3 & 8.8 & 0.037 & 5.0--7.5 & 0.11--0.16 & 3.1 & 424 \\
\end{tabular}
\end{ruledtabular}
\end{table*}

\section*{Methods}

\subsection*{Growth and samples}
The quantum wells were grown by molecular-beam epitaxy, arsenic-capped, and the cap was desorbed in an atomic-hydrogen atmosphere under ultra-high vacuum before metal deposition. Al, Sn, V, Nb, Ta and Re were deposited by molecular-beam evaporation at a substrate temperature of 7~K, which suppresses interdiffusion and compound formation at the interface. A nominal 2.5-nm Al protection layer was oxidized in situ at low temperature immediately after deposition, before the sample was warmed (except for the Al film). Every sample was subsequently warmed to room temperature and cooled again for measurement. The cap forms no active superconducting layer. A 5-nm Ta film retains $T_c=1.6$~K, below the value of the Al sample, and Re shows an increasing $T_c$ with decreasing thickness as the amorphous phase is stabilized, neither of which an Al-dominated transition could reproduce.

\subsection*{Magnetotransport}
Four-terminal resistance was measured on unpatterned material with collinear probes 1~mm apart, in perpendicular and tilted fields up to 14~T at temperatures of 1.8--20~K. The measured four-terminal resistance is proportional to the longitudinal resistivity of the stack, $R=g\rho_{xx}$ with $g$ a geometric factor. The uncapped wells settle this: in the quantum Hall regime their resistance falls to $1$--$5\%$ of its zero-field value, with the deep minima at integer filling factor, whereas $g/\sigma_{xx}$ would diverge there and would place maxima at the same fields. Only temperatures at which the film is in its normal state over the fit window were used: $T\geq 7 $~K for 3-nm Re, $T\geq5$~K for Nb and 1-nm Re, $T\geq3$~K for 6-nm Re and $T\geq2$~K otherwise. Each fit window begins above the upper critical field of its film at the analysis temperature, at $0.50$~T for Al, Sn and V, $0.60$~T for Nb and $0.75$~T for Ta and all three Re films, and at $0.75$--$1.00$~T for the interface branches.

\subsection*{Frequencies and densities}
Oscillation frequencies were extracted from the polynomial-detrended resistance against $1/B$ with a chirp-$z$ transform. Densities follow from $n=2eF/h$ for an unresolved spin-degenerate orbit. The instantaneous frequency of the interface branches drifts by less than 4\% across the fit window, except for the V interface branch, which is the weakest in the series and whose envelope shows a minimum near 3.7~T, so no lifetime is quoted for it. No beat is resolved in any fundamental branch. A spin-split pair would place its first envelope node at $B=2\Delta F$, and the lowest field in any fundamental window is $0.50$--$0.75$~T, so $\Delta F<0.25$--$0.38$~T, corresponding to $\Delta n<6.6\times10^{9}$~cm$^{-2}$ and a Rashba coefficient below $3$~meV\,nm for the buried well. The single-component form of Eq.~\eqref{eq_shunt_model} therefore applies. The wells of Ref.~\cite{Zimmerman2026} beat at $\Delta F\simeq1$~T. At a common chemical potential the occupation energy of a two-dimensional subband is
\begin{equation}
\EF-E_i=\frac{\pi\hbar^2n_i}{m_i},
\label{eq_subband_occupation}
\end{equation}
in the parabolic limit. InAs is strongly nonparabolic, so we solve the two-band form $E(1+E/E_g')=\hbar^2k^2/2m_0^*$ with the measured cyclotron mass identified as $m^*=m_0^*(1+2E/E_g')$ at $\EF$ and $E_g'=0.45$~eV, which raises the occupations by $3\%$ for the buried well, $8\%$ for the refractory interface branches and $16\%$ for Al and Sn. This places the fundamental state $14$--$16$~meV below $\EF$ and the interface subband $33$--$40$~meV (refractory metals) or $85$--$99$~meV (Al and Sn) below $\EF$, that is $18$--$26$~meV and $70$--$84$~meV below the fundamental state respectively. The values change by less than 2~meV over $E_g'=0.40$--$0.50$~eV.

\subsection*{Cyclotron masses}
Masses were obtained from the Lifshitz--Kosevich temperature dependence of the amplitude in sliding windows of $2.5$ periods, averaged over the window in which the oscillation contains a single frequency. Traces in which the film was superconducting over part of the window were excluded, which changes the 1-nm Re fundamental mass by about 10\% relative to an analysis that includes them. No mass is available for 3-nm Re, whose film is normal only above 7~K.

\subsection*{Quantum lifetimes beneath a shunt}
The Lifshitz--Kosevich amplitude of one subband factorizes as $A=A_0R_TR_D$ with $R_T=X/\sinh X$, $X=2\pi^2\kb T/\hbar\omega_c$ and $R_D=e^{-\pi/\omega_c\tau_q}$, and the oscillation is fundamentally one of the scattering rate, $1/\tau(B)=(1/\tau_0)[1+\delta(B)]$ with $\delta\propto R_TR_D\cos(2\pi F/B+\varphi)$. For a single channel $\rho_{xx}=m^*/(ne^2\tau)$ exactly, so $\delta\rho_{xx}/\rho_0=\delta$ and the conventional Dingle plot of $\ln(A\sinh X/X)$ against $1/B$ applies~\cite{Coleridge1989,Coleridge1991}. Differentiating the Drude tensor with respect to $1/\tau$ gives $\delta\sigma_{xx}=\sigma_0P\,\delta$ and $\delta\sigma_{xy}=\sigma_0Q\,\delta$, with $P=(\mu_t^2B^2-1)/(1+\mu_t^2B^2)^2$ and $Q=-2\mu_tB/(1+\mu_t^2B^2)^2$, so that
\begin{equation}
\delta\rho_{xx}=\frac{(\sigma_{xy}^2-\sigma_{xx}^2)\,\delta\sigma_{xx}-2\sigma_{xx}\sigma_{xy}\,\delta\sigma_{xy}}{(\sigma_{xx}^2+\sigma_{xy}^2)^2},
\label{eq_tensor}
\end{equation}
summed over channels. For a single channel the numerator collapses through $(\mu^2B^2-1)^2+4\mu^2B^2=(1+\mu^2B^2)^2$ to give $\delta\rho_{xx}/\rho_0=\delta$ identically, which is why the uncapped wells need no correction. With a metal film in parallel the total $\sigma_{xx}$ is dominated by the film and the total $\sigma_{xy}$ is that of the semiconductor alone, and Eq.~\eqref{eq_tensor} reduces to $\delta R=-(\Rsm^2/g)\,\delta\sigma_{xx}$, which gives Eq.~\eqref{eq_shunt_model}. This is the $\rho_{xx}$ form of the construction of Ref.~\cite{Candido2023} with a metallic channel added. We evaluate Eq.~\eqref{eq_tensor} in full for every branch. The $\sigma_{xy}$ terms shift $\mu_q$ by less than 4\% at the $\mu_t\rightarrow\infty$ end of each interval and by up to a factor of two at the $\tau_t/\tau_q=2$ end, where the node $\mu_tB=1$ enters the window, and the intervals quoted include them. The prefactor is the field-dependent envelope that Ref.~\cite{Zimmerman2026} attributes to current redistribution into the metal.

Fits used the amplitude at the lowest normal-state temperature, bins with filling factor $\nu\geq4$ and amplitude above four times the empty-band noise floor, and $\Rsm(B)$ from the 20~K trace. Because $\mu_t$ is not identified by the smooth background, which the film resistivity, the geometric factor and the two subband mobilities fit equally well over a wide range (SI), $\mu_q$ is reported as an interval from its $\mu_t\rightarrow\infty$ limit, in which $|P|\rightarrow(\mu_tB)^{-2}$, to the fixed point $\mu_t=2\mu_q$. The uncapped wells give $\tau_t/\tau_q=2$--$4$, from a zero-field sheet resistance corresponding to $\mu_t=(2$--$4)\times10^4$~cm$^2$V$^{-1}$s$^{-1}$ against $\mu_q=(1.0$--$1.2)\times10^4$. Screening by the film and isotropic escape into it lower the ratio towards unity, so the hybrid ratio lies between about 1 and the uncapped value, and $\mu_q$ is reported over that range. A sweep of $\tau_t/\tau_q$ over $[1,4]$ for every branch is given in the SI. The intervals reproduce to $1$--$10\%$ across analysis temperatures of $2$--$6$~K, whereas the conventional Dingle plot of the same data returns values that move by factors of $1.5$--$4$ with temperature, because it omits the shunt prefactor, whose contribution to the apparent slope changes as the thermal factor re-weights the window. The Sn interface branch is fitted at $B\geq1$~T only. A one-subband parallel-conduction model of the background is not an alternative, since it returns $\tau_t<\tau_q$ once the interface branch is occupied.

\subsection*{Spin zeros}
In a tilted field the orbital quantization is set by $B_\perp=B\cos\theta$ whereas the Zeeman energy depends on the total field, so $F(\theta)=F(0)/\cos\theta$ and the spin factor crosses zero when
\begin{equation}
\frac{|g_{\mathrm{eff}}|m^*}{\me}=\cos\theta_z .
\label{eq_spinzero}
\end{equation}
Amplitudes against tilt were fitted with a coincidence lineshape damped by $\exp(-d\tan^2\theta)$, and the null was confirmed model-free by the inversion of the band-passed waveforms either side of it. The spin zero depends only on the Zeeman-to-cyclotron ratio at fixed $B_\perp$ and is unaffected by the amplitude transfer function of Eq.~\eqref{eq_shunt_model}, which cancels between angles.

\subsection*{Coupling bounds}
In the local wide-band approximation, integrating out the superconductor gives the retarded self-energy
\begin{equation}
\Sigma^{R}(\omega)=-\Gamma\,\frac{\omega+\Delta_0\tau_x}{\sqrt{\Delta_0^2-(\omega+i0^+)^2}},
\label{eq_selfenergy}
\end{equation}
with $\Gamma$ the normal-state hybridization rate~\cite{vanHeck,Reeg2018,Zimmerman2026}. For $|\omega|\ll\Delta_0$ the quasiparticle weight is
\begin{equation}
Z=\left(1+\frac{\Gamma}{\Delta_0}\right)^{-1},
\label{eq_Z}
\end{equation}
and the low-energy terms are renormalized as
\begin{equation}
\begin{aligned}
\msub &= \frac{\mN}{Z}, \\
g_{\mathrm{sub}} &= Z\gSm+(1-Z)\gm , \\
\alpha_{\mathrm{sub}} &= Z\alpha ,
\end{aligned}
\label{eq_renorm}
\end{equation}
with the $g$-factors signed in a common convention and $\alpha$ the Rashba coupling~\cite{Reeg2018}. Taking $\Delta_0\rightarrow0$ gives $\Sigma^R=-i\Gamma$, which enters the Dingle factor but has no real part and therefore leaves the cyclotron spacing and the Zeeman-to-cyclotron ratio unchanged.

Escape into the metal is a golden-rule process,
\begin{equation}
\frac{2\Gamma_i}{\hbar}=\frac{2\pi}{\hbar}\sum_{\nu}\left|\langle\nu|H_{\mathrm T}|\psi_i\rangle\right|^2\delta(E-E_\nu)=\frac{2\pi}{\hbar}\,|t_i|^2\nu_{\mathrm m},
\label{eq_golden_rule}
\end{equation}
with $|t_i|^2\propto|\psi_i(z_{\mathrm m})|^2$ the envelope weight at the boundary~\cite{Bardeen1961,Antipov2018,Mikkelsen2018}. Rates of independent channels add, so
\begin{equation}
\frac{1}{\tau_q}=\frac{1}{\tau_{q,\mathrm{dis}}}+\frac{2\Gamma}{\hbar},
\label{eq_lifetime_decomposition}
\end{equation}
and since the disorder rate cannot be negative,
\begin{equation}
\Gamma\leq\frac{\hbar}{2\tau_q}=\frac{\hbar e}{2m^*\mu_q}.
\label{eq_gamma_bound}
\end{equation}
For the buried well the uncapped samples measure $\tau_{q,\mathrm{dis}}$ directly, and subtracting Eq.~\eqref{eq_lifetime_decomposition} for the hybrid and the reference gives
\begin{equation}
\Gamma_{\mathrm{QW}}\leq\frac{\hbar}{2}\left[\frac{1}{\tau_q^{\mathrm{hyb}}}-\frac{1}{\tau_q^{\mathrm{bare}}}\right],
\label{eq_gamma_differential}
\end{equation}
valid whenever metallization does not reduce the disorder seen by the well, evaluated with the shortest hybrid lifetime and the longest uncapped one. For a state confined against the metal over a length $L$ with velocity $v_z$, Eq.~\eqref{eq_golden_rule} reduces to $\Gamma=\hbar v_z\mathcal{T}/4L$, with $\mathcal{T}$ the transmission per encounter and $v_z/2L$ the attempt frequency. A triangular accumulation well with the measured interface densities gives $\hbar v_z/4L\simeq17$--$30$~meV. Induced-gap ceilings solve the pole of Eq.~\eqref{eq_selfenergy}, $E\bigl(\sqrt{\Delta_0^2-E^2}+\Gamma\bigr)=\Gamma\Delta_0$, at each coupling bound. In the low-energy limit this reduces to
\begin{equation}
\Deltaind/\Delta_0\simeq1-Z .
\label{eq_gapZ}
\end{equation}

\subsection*{Data availability}
Full analysis script getting from the raw data to figures is deposited at the public repository and will be available at the time of manuscript submission. 

\begin{acknowledgments}
The authors acknowledge support from the Microsoft Quantum Pioneer Program. G.P.M. thanks the Robert Mehrabian College of Engineering and the Department of Materials at the University of California, Santa Barbara, for support. We also acknowledge the U.S.\ Department of Energy under Award No.\ DE-SC0025017 for cryogenic-temperature superconductor synthesis, and the University of California, Santa Barbara (UCSB) National Science Foundation (NSF) Quantum Foundry through the Q-AMASE-i Program via Award No.\ DMR-1906325 for the LTMBE facility and for Graduate Student Fellowship support (T.A.J.v.S.). We further acknowledge the use of shared facilities of the UCSB MRSEC (NSF DMR-2308708) and the UCSB Nanofabrication Facility. G.P.M. thanks Leo Kouwenhoven for helpful discussion.
\end{acknowledgments}

\end{document}